\documentclass[12pt]{article}
\usepackage{authblk}
\usepackage[bookmarksnumbered, colorlinks, plainpages]{hyperref}
\usepackage{amsmath, amsthm, amscd, amsfonts, amssymb, graphicx, color, booktabs}
\numberwithin{equation}{section}
\definecolor{email}{rgb}{0.00,0.00,0.84}
\begin{document}
\setcounter{page}{1}

\title{\large \bf 12th Workshop on the CKM Unitarity Triangle\\ Santiago de Compostela, 18-22 September 2023 \\ \vspace{0.3cm}
\LARGE The status of New Physics in $b \to c \ell\nu$ transitions}

\author[1,2]{Marco Fedele\footnote{marco.fedele@ific.uv.es}}
\affil[1]{Institute for Theoretical Particle Physics, Karlsruhe Institute of Technology (KIT), Wolfgang-Gaede-Str. 1, D-76131 Karlsruhe, Germany}
\affil[2]{Departament de Física Teòrica, IFIC, Universitat de València – CSIC,
Parque Científico, Catedrático José Beltrán 2, E-46980 Paterna, Spain}
\maketitle

\begin{abstract}
Motivated by the recent theoretical and experimental developments in $b \to c \ell\nu$ transitions, the aim of these proceedings is to critically analyse the current situation in this sector. In particular, we will inspect the possibility of explaining the current anomalous pattern by means of New Physics effects coupled with light charged leptons, motivated both by the measurement of the LFUV ratio ${\cal R}(\Lambda_c)$ and by the recent application of the Dispersive Matrix approach to estimate Form Factors in $B \to D^*$ transitions. These proceedings are based on Refs.~\cite{Fedele:2022iib,Fedele:2023ewe}.
\end{abstract} \maketitle

\section{Introduction}
The Standard Model (SM) of particle physics has been extremely successful in describing the behavior of fundamental particles and their interactions. However, it is widely believed that there must be additional, yet unknown, physics beyond the SM. A promising approach to search for new physics is to look for the violation of symmetries of the SM, such as lepton flavor universality (LFU), which is only broken in the SM Lagrangian by the small Yukawa couplings.

Over the past few years, several hints for the violation of LFU have emerged. In particular, the ratios of the semi-leptonic $B$-meson decays ${\cal R}(D^{(*)})\equiv {\rm BR}(B\to D^{(*)} \tau \bar\nu)/{\rm BR}(B\to D^{(*)} \ell \bar\nu)$, with $\ell=e, \mu$, show deviations from the SM predictions which can be interpreted in terms of an over-abundance of taus, with a tension with the recent SM predictions at the level of $3.2\,\sigma$~\cite{HFLAV:2022pwe}. The recent measurement of ${\cal R}(\Lambda_c)\equiv {\rm BR}(\Lambda_b \to\Lambda_c \tau\bar\nu)/{\rm BR}(\Lambda_b \to\Lambda_c \ell\bar\nu)$~\cite{LHCb:2022piu}, on the other hand, does not point towards a strong tension with the SM, actually hinting to an under-abundance of taus in contrast to what assumed when new physics (NP) effects couple to taus~\cite{Blanke:2018yud,Blanke:2019qrx}. A possible explanation could be given assuming NP effects coupled to light leptons, an option viable when employing the Dispersive Matrix (DM) approach in the Form Factors (FFs) computation, which reduces the tension in $\mathcal{R}(D^{*})$ to $1.3\,\sigma$~\cite{Martinelli:2021myh}. However, when employing this approach, it is fundamental to confront the predictions for the $D^*$ longitudinal polarization fraction $F_L^{\ell}$ and the forward-backward asymmetry $A_{\rm FB}^\ell$ with their measurements, which have been recently observed with great precision for light charged leptons~\cite{Belle:2023bwv,BelleIISem}.

The goal of this study was therefore to critically scrutinize the compatibility of data. Our aim was to understand if, on the one hand, a compatibility in the data concerning the three LFUV ratios can be found in the SM or beyond, and, on the other hand, if the employment of DM FFs increase the overall theoretical description of data.

\section{The Effective Hamiltonian}

We perform our analyses by means of the effective Hamiltonian
\begin{equation}
 {\cal H}_{\rm eff}=  2\sqrt{2} G_{F} V^{}_{cb} \big[(1+C_{V_L}^l) O_{V_L}^l + C_{S_L}^l O_{S_L}^l +C_{S_R}^l O_{S_R}^l+C_{T}^l O_{T}^l\big] \,,
\label{Heff}
\end{equation}
where $l=e,\mu,\tau$ and
\begin{equation}
\renewcommand{\arraystretch}{1.8}
\begin{array}{l}
   O_{V_L}^l  = \left(\bar c\gamma ^{\mu } P_L b\right)  \left(\bar l \gamma_{\mu } P_L \nu_{l}\right)\,, \qquad\qquad O_{S_L}^l  = \left( \bar c P_L b \right) \left( \bar l P_L \nu_{l}\right)\,,   \\
   O_{S_R}^l  = \left( \bar c P_R b \right) \left( \bar l P_L \nu_{l}\right)\,, \qquad\qquad O_{T}^l  = \left( \bar c \sigma^{\mu\nu}P_L  b \right) \left( \bar l \sigma_{\mu\nu} P_L \nu_{l}\right)\,.
\end{array}
\label{Oeff}
\end{equation}
where the Wilson coefficients (WCs) $C^l_i$ describe a genuine NP effect, and vanish in the SM. Our analysis is performed for a heavy NP scale of 2$\,$TeV. We therefore connect our results to the decay scale $\mu=\mu_b = 4.2\,$GeV by means of renormalization-group evolution (RGE), which implies
\begin{equation}
\renewcommand{\arraystretch}{1.8}
\begin{array}{l}
C_{V_L}^l (\mu_b) = 1.12 \, C_{V_L}^l (2\,\mbox{TeV})\,, \label{wcrun}\\[1mm]
C_{S_R}^l(\mu_b) =2.00\, C_{S_R}^l (2\,\mbox{TeV}) \,, \\[1mm]
  \left( \begin{array}{c}
      C_{S_L}^l(\mu_b) \\ C_T^l(\mu_b)           
   \end{array}\right) =  \left( \begin{array}{rr}
          1.91 & -0.38 \\
           0.    & 0.89 
   \end{array}\right)
  \left( \begin{array}{c}
         C_{S_L}^l(2\,\mbox{TeV}) \\ C_T^l(2\,\mbox{TeV})    
   \end{array}\right) . 
\end{array}
\end{equation}

\section{NP scenarios}

Let's briefly summarize here the NP scenarios that could be interesting to consider in this context.
\begin{itemize}
\item Scalar Leptoquarks
\begin{itemize}
\item $S_1$, an $SU(2)_L$-singlet scalar that at the low scale contributes to $C_{V_L}^l$ and/or the combination  $C_{S_L}^l (\mu_b)\simeq - 8.9 C_T^l (\mu_b)$. 
\item $R_2$, a weak doublet scalar whose footprints are described by a contribution satisfying $C_{S_L}^l (\mu_b) \simeq 8.4 C_T^l (\mu_b)$ at the low scale. 
\item $S_3$, an $SU(2)_L$-triplet scalar that is parametrized at the low scale by the WC $C_{V_L}^l$.
\end{itemize}
\end{itemize}

\begin{itemize}
\item Vector Leptoquarks
\begin{itemize}
\item $U_1$, an $SU(2)_L$-singlet vector that produces at the low scale the WCs $C_{V_L}^l$ and/or $C_{S_R}^l$.
\item $V_2$, a weak doublet vector whose effects at the decay scale can be described by means of the WC $C_{S_R}^l$.
\end{itemize}
\end{itemize}

\begin{itemize}
\item Charged Bosons
\begin{itemize}
\item $H^\pm$, A charged scalar boson ($H^\pm$) generates the WCs $C_{S_R}^l$ and $C_{S_L}^l$.
\item $W'$, a new charged vector boson which generates $C_{V_L}^l$.
\end{itemize}
\end{itemize}

\section{Compatibility of the LFUV ratios}

We discuss here our results concerning the compatibility of LFUV data~\cite{Fedele:2022iib}. We start by performing a fit to the LFUV ratios only, and in the case where we can satisfactorily reproduce the three of them we consider additional constraints, like, e.g., the $D^{*-}$ polarization~\cite{Belle:2019ewo}.

\begin{figure*}[!t!]
\centering
\includegraphics[width =0.65\textwidth]{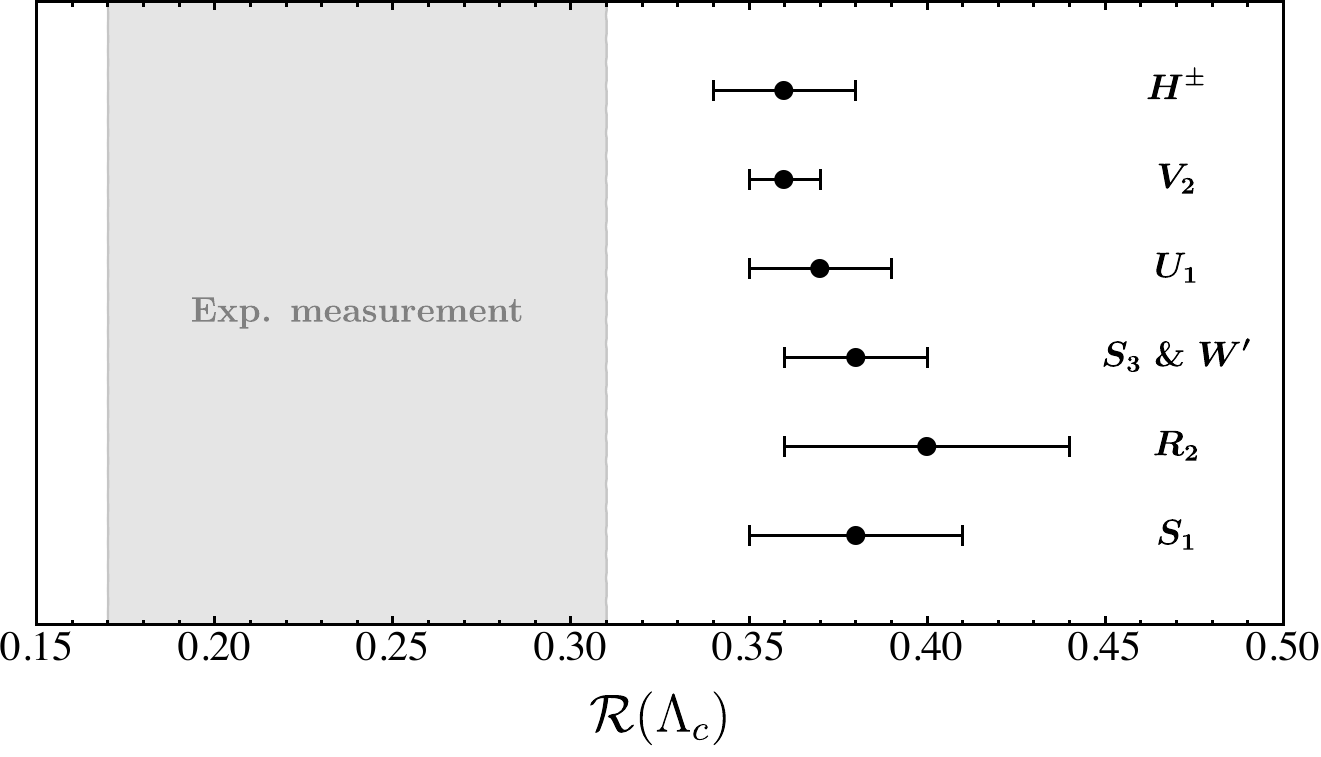}
\caption{Predicted values for ${\cal R}(\Lambda_c)$ when fitting ${\cal R}(D^{(*)})$ in different NP models.
}
\label{fig:RLpred}
\end{figure*}

We first observe that none of the models discussed in the previous section are capable of describing these ratios in a satisfactory way, assuming that NP couples to taus only. To further understand the level of the discrepancy, for each scenario capable of fitting ${\cal R}(D^{(*)})$ we predicted the value for ${\cal R}(\Lambda_c)$, and vice versa. When predicting the value of ${\cal R}(\Lambda_c)$ from the mesons LFU ratios, a large prediction for the baryon ratio was obtained, compatible with the prediction of the sum rule and $\sim 2\sigma$ above its measured value, see Fig.~\ref{fig:RLpred}. On the other hand, when predicting the values of ${\cal R}(D^{(*)})$ from a fit to the baryon ratio, a value for the latter ratio complying with data would imply values for the former ratios $\sim 2\sigma$ below their measured values.

We proceed inspecting also the cases where 2 NP fields are allowed, coupled one to taus and the other to light leptons, respectively. Out of all the possible configurations, only two scenarios have been found capable to reproduce all the LFUV ratios in a satisfactory way. The first viable model is composed by the pair formed by $S_1^\ell$ and $R_2^\tau$, while the second one consists in the pair formed by $S_1^\ell$ and $H^{\pm\tau}$. These models appear viable due to the presence of $S_1^\ell$ in both scenarios, coming with a strong contribution to the scalar and tensor currents, found to be equal to $C_{S_L}^\ell=-8.9C_T^\ell \simeq \pm 1$, together with a strong cancellation of the SM component, induced by $C_{V_L}^\ell\simeq -1$. These values for the WCs are however not phenomenologically viable, when the above additional constraints are taken into account. Indeed, high-$p_{\rm T}$ lepton tail searches constrain the vector coupling to $|C_{V_L}^{e}|<0.25$ \cite{Iguro:2020keo,Allwicher:2022mcg}, with the $B\to K^*\nu\bar{\nu}$ measurement further implying $-0.011\le C_{V_L}^\ell\le 0.027$~\cite{Endo:2021lhi}. The tensor component is also strongly constraint for light leptons due to high-$p_{\rm T}$ searches~\cite{ATLAS:2019lsy}, implying $|C_T^{e}|< 0.32$, or by analysis of angular distribution~\cite{Belle:2017rcc,Belle:2018ezy} and $D^{*-}$ polarization data~\cite{Belle:2019ewo}, which requiring $|C_T^\ell|\le 0.05$~\cite{Jung:2018lfu,Iguro:2020cpg}. Similarly, a crude estimate that comes from Unitarity Triangle Analysis bounds these WCs to $|C_{V_L}^{\ell}| < 0.025$ and $|C_T^{\ell}| < 0.25$.

We therefore concluded that it is not possible to address in a satisfactory way present data concerning LFUV ratios, as current experimental results for $\mathcal{R}(D^{(*)})$ and $\mathcal{R}(\Lambda_c)$ show an inconsistent pattern of behaviour and deviations.

\section{The impact of $F_L^{\ell}$ and $A_{\rm FB}^\ell$ on Form Factors determinations}

\begin{figure*}[!t!]
\centering
\includegraphics[width =0.19\textwidth]{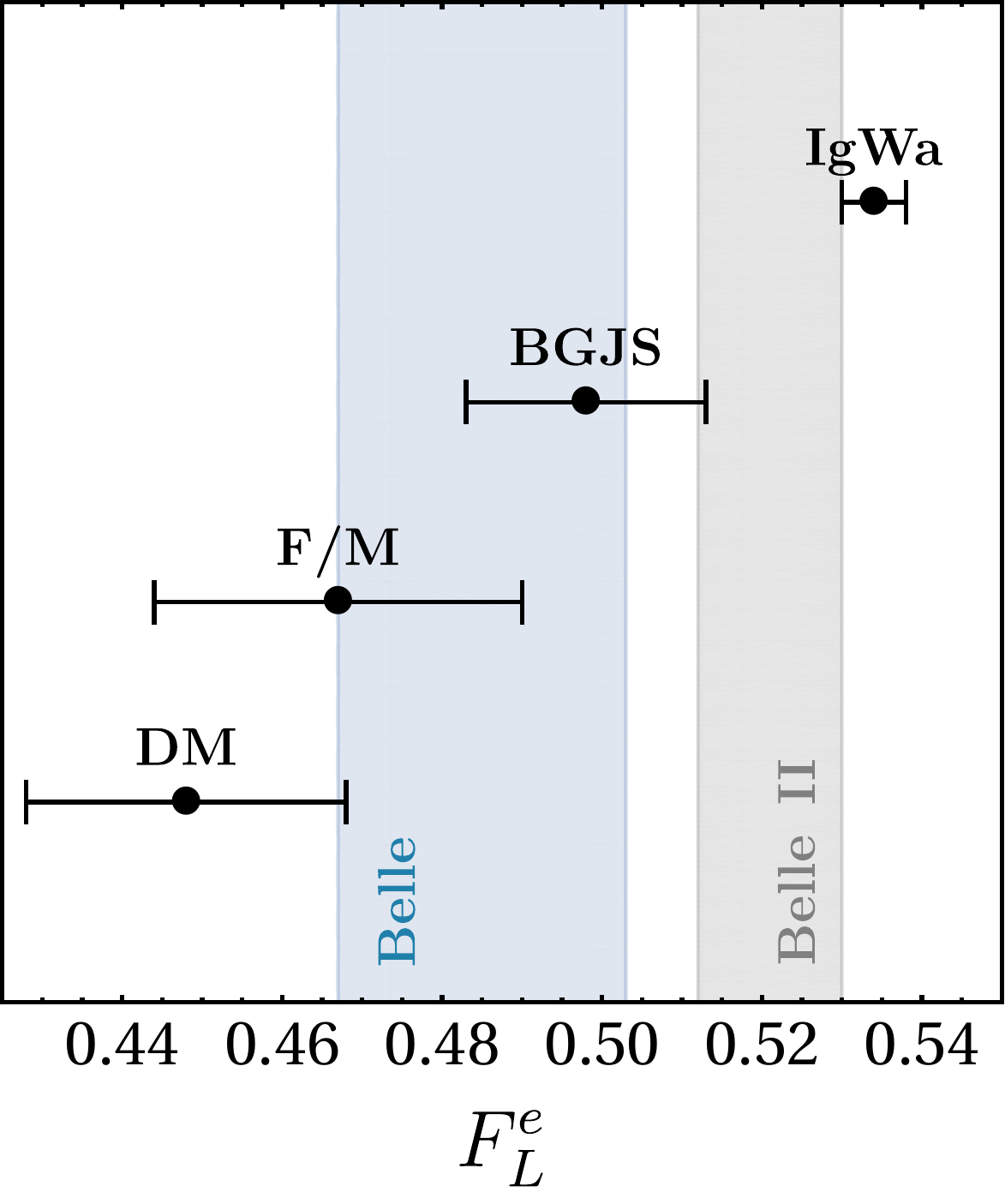}
\includegraphics[width =0.19\textwidth]{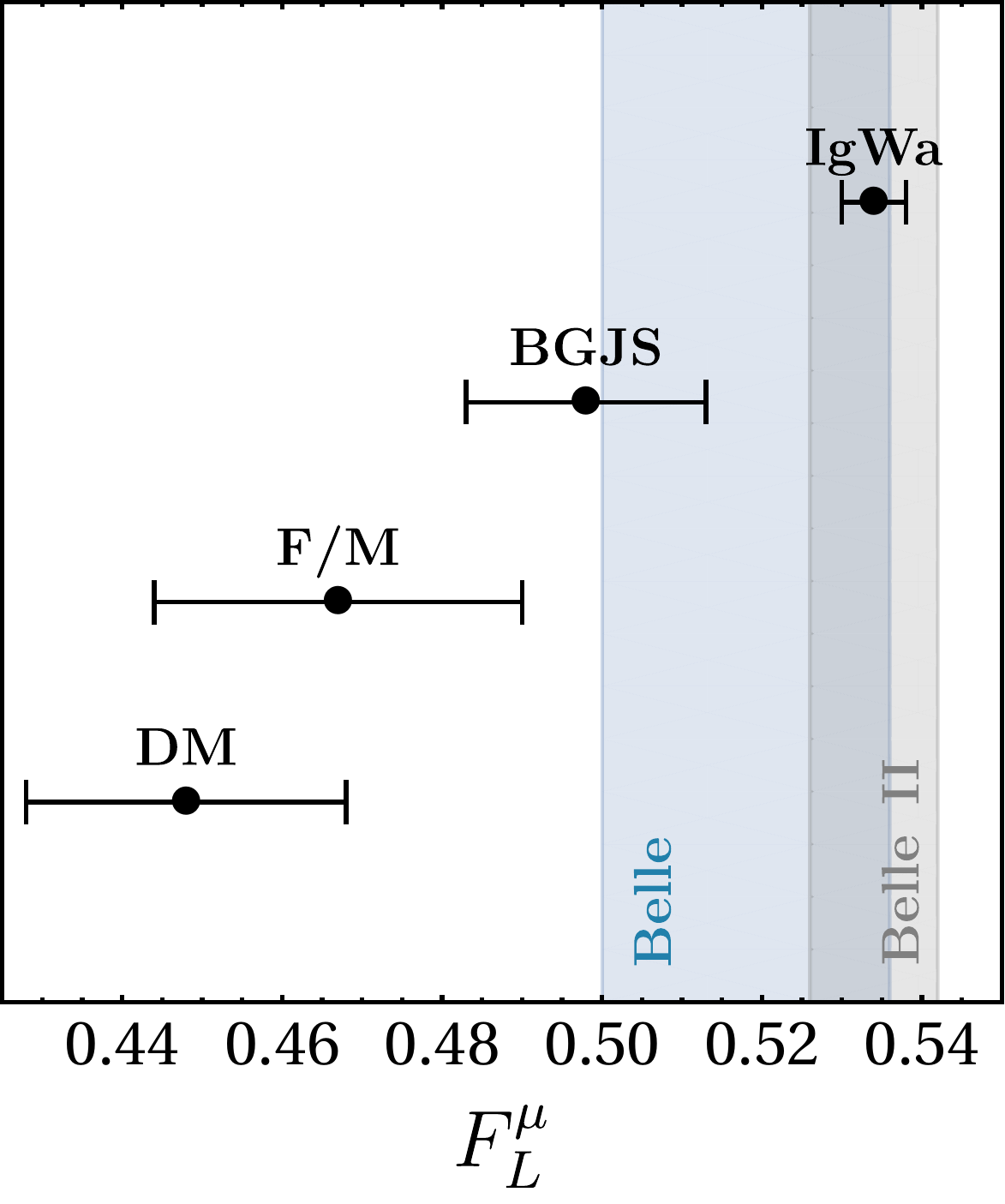}
\includegraphics[width =0.19\textwidth]{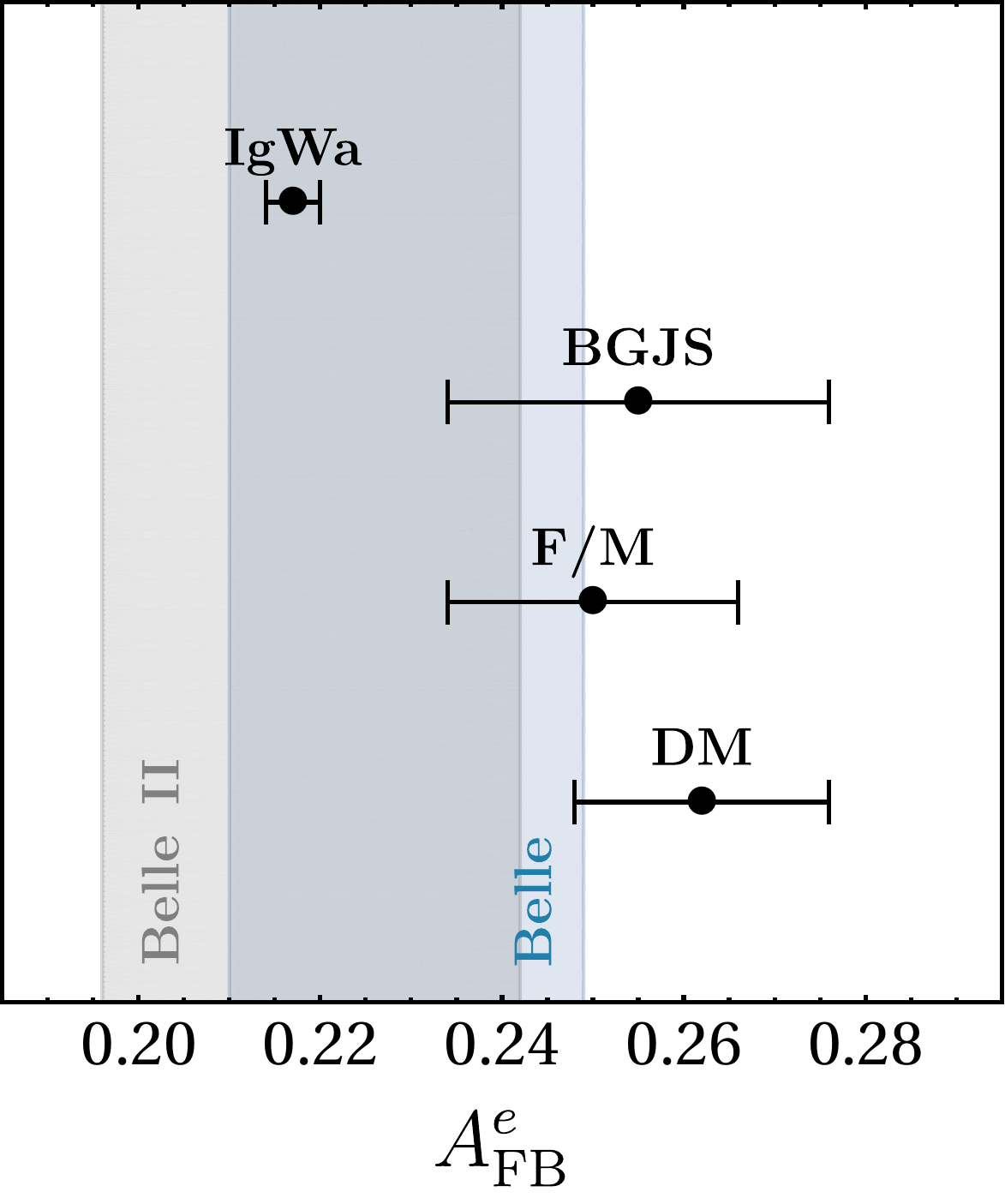}
\includegraphics[width =0.19\textwidth]{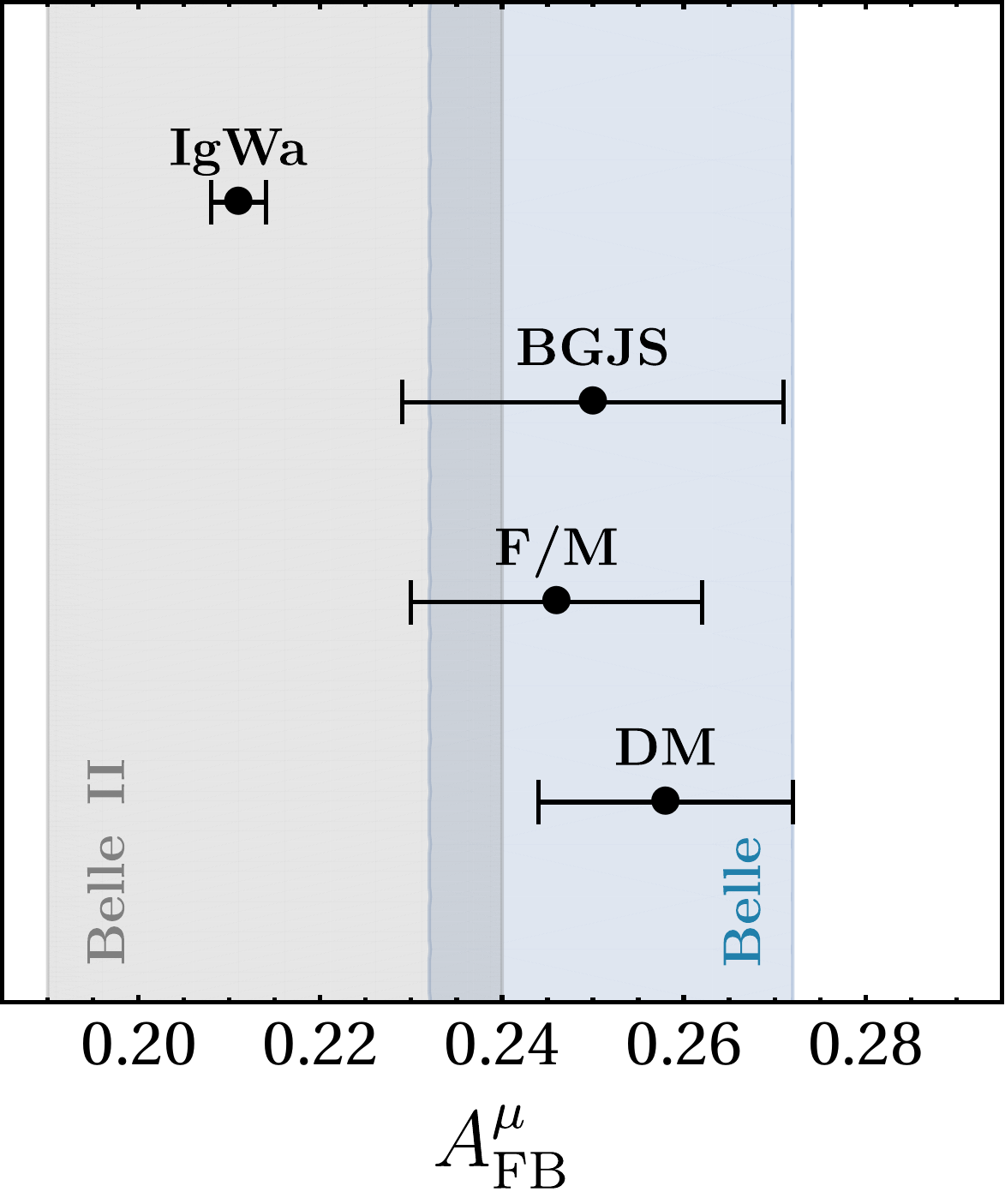}
\includegraphics[width =0.19\textwidth]{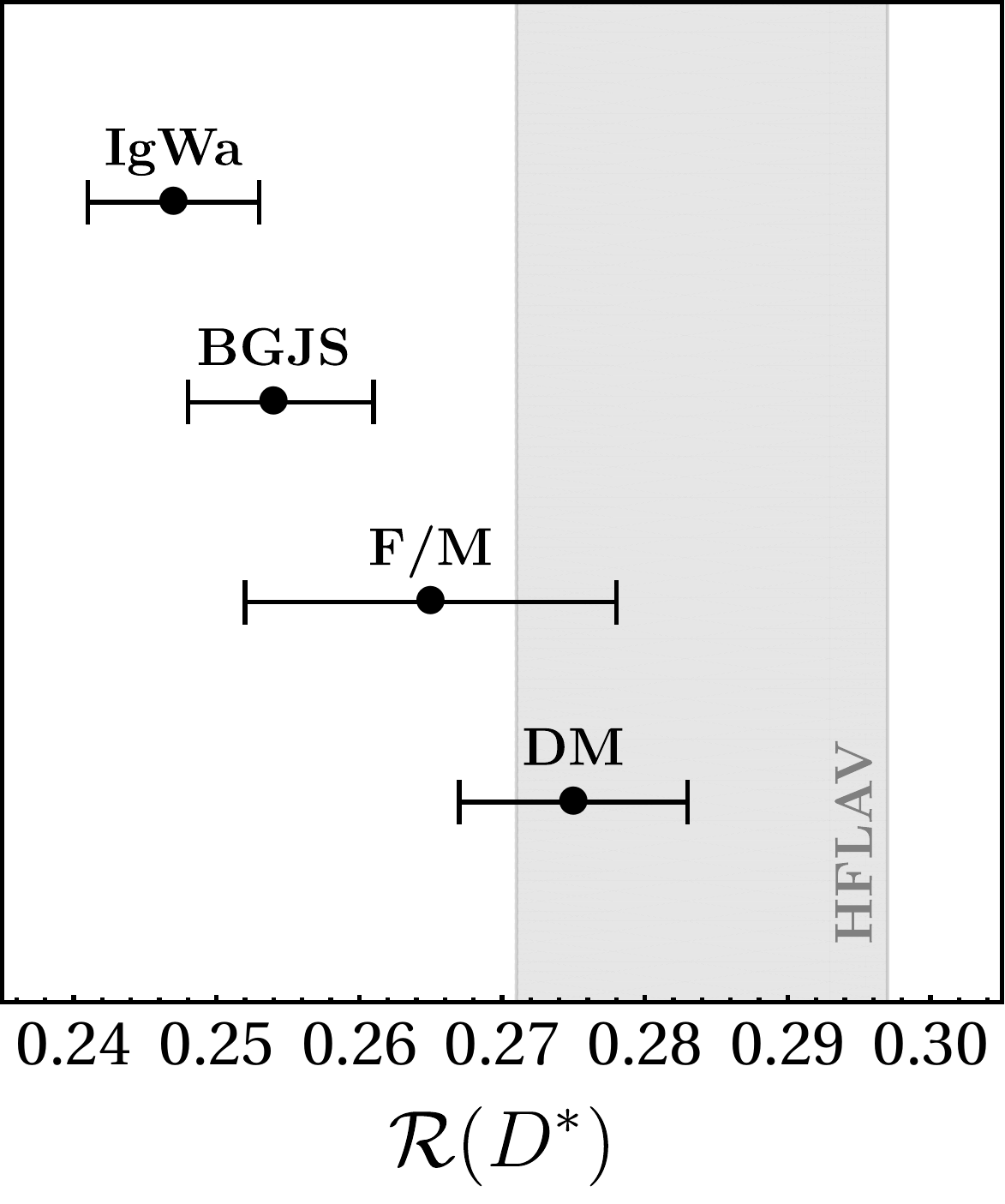}
\caption{Comparisons among measurements and SM predictions for $F_L^{\ell}$, $A_{\rm FB}^\ell$ and $R(D^*)$ at the $1\,\sigma$ level for different FFs estimates.
}
\label{fig:comparison}
\end{figure*}

\begin{figure*}[!t!]
\centering
\includegraphics[width =0.65\textwidth]{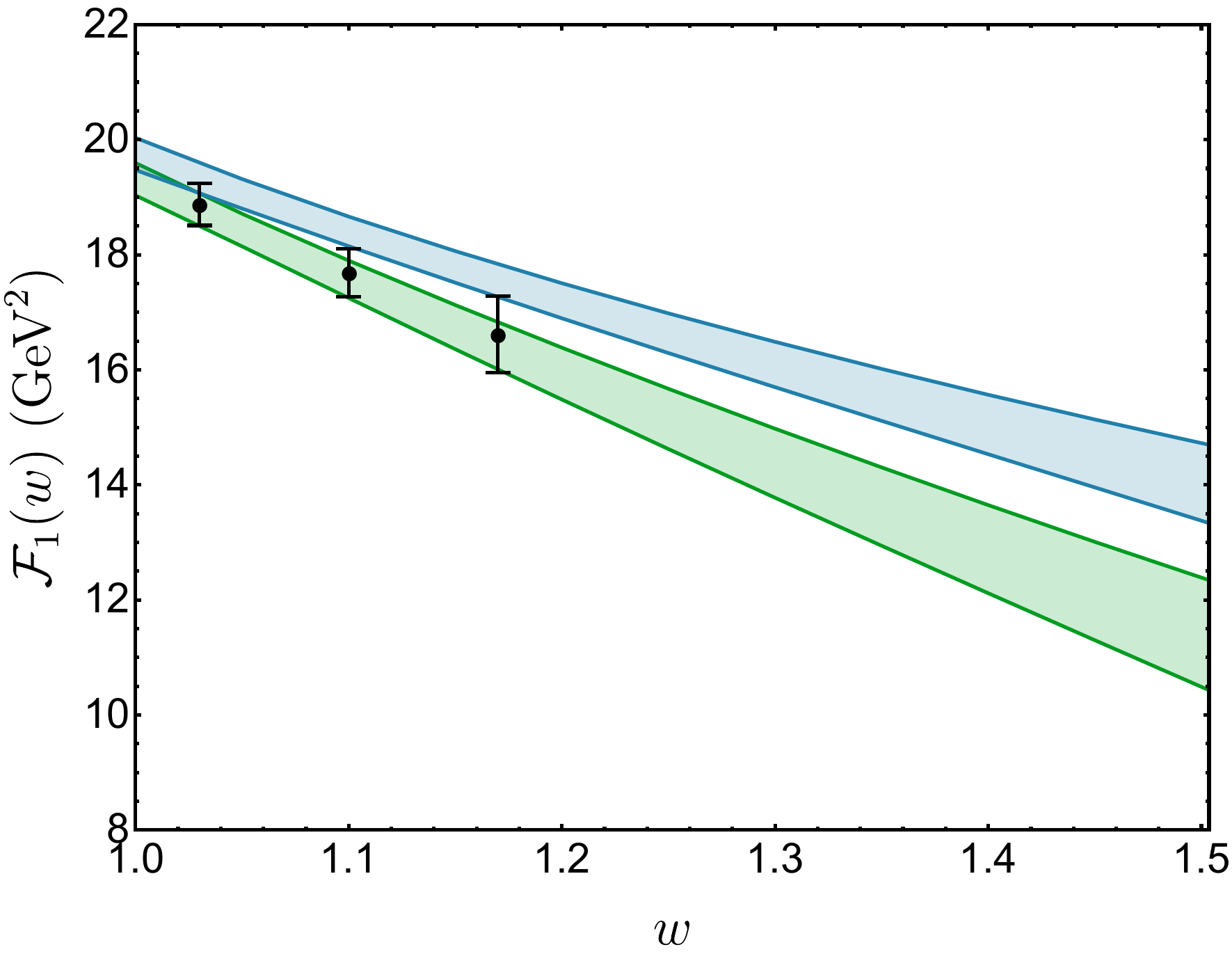}
\caption{Green band: DM estimate~\cite{Martinelli:2021myh}. Blue band: modified FF after performing a fit to data from Refs.~\cite{HFLAV:2022pwe,Belle:2023bwv,BelleIISem}. Black data points: F/M input values~\cite{FermilabLattice:2021cdg}.
}
\label{fig:F1}
\end{figure*}

We conclude these proceedings reviewing the connection between FFs determinations and their predictions for $R(D^*)$, $F_L^{\ell}$ and $A_{\rm FB}^\ell$ observables, where light charged leptons are considered in the final state~\cite{Fedele:2023ewe}. In particular, we consider four different approaches: IgWa, an approach based on Heavy Quark Effective Theory~\cite{Iguro:2020cpg}; BGJS, an estimate based on the BGL parametrization~\cite{Bigi:2017njr,Bigi:2017jbd,Gambino:2019sif}; F/M, the first lattice QCD estimate to go beyond zero-recoil~\cite{FermilabLattice:2021cdg}; and DM, the Dispersive Matrix approach that uses the F/M lattice results as an input, further filtered by unitarity~\cite{Martinelli:2021myh}. The starting point of this analysis was the observation that, according to the employed FFs determination, a reduced discrepancy in $R(D^*)$ is counterbalanced by an increased one in $F_L^{\ell}$ and $A_{\rm FB}^\ell$, see Fig.~\ref{fig:comparison} where the theory predictions are confronted with data~\cite{HFLAV:2022pwe,Belle:2023bwv,BelleIISem}. Moreover, if one performs a fit to data form Refs.~\cite{HFLAV:2022pwe,Belle:2023bwv,BelleIISem} in the DM approach, rather than predicting these quantities, the result is a shift in the FFs (mainly in the so-called $\mathcal{F}_1$ one) to values closer to the ones estimated by the IgWa or BGJS approaches, reducing the tension in $F_L^{\ell}$ and $A_{\rm FB}^\ell$ but reintroducing the one in $R(D^*)$; moreover, the new shape of $\mathcal{F}_1$ is such that it is not even compatible with the input F/M values, see Fig.~\ref{fig:F1}. Hence, we are in a situation in which, again, it is not currently possible to reproduce all experimental at the same time in the SM. 

Therefore, our next step is to study whether the DM approach, which is the one better reproducing $R(D^*)$ but with the largest deviations in $F_L^{\ell}$ and $A_{\rm FB}^\ell$, can be recovered by allowing NP effects equally coupled to light leptons. Indeed, such an option is not viable for the other approaches, where the amount of required NP coupled to light leptons and the consequent extracted value for $|V_{cb}|$ from this channel would be in strong contrast with CKM unitarity fits. To this end, we performed once again a fit to data from Refs.~\cite{HFLAV:2022pwe,Belle:2023bwv,BelleIISem}, allowing now also for NP effects. The bounds at the $1\,\sigma$ level read
\begin{equation}
\renewcommand{\arraystretch}{1.8}
\begin{array}{l}
   C_{V_L}^l  = -0.054\pm0.015\,, \qquad\quad\; C_{S_L}^l  \in [-0.07,0.02]\,,   \\
   C_{S_R}^l \in [-0.05,0.03] \,, \qquad\qquad\quad C_{T}^l  \in [-0.01,0.02]\,.
\end{array}
\end{equation}
$C_{V_L}$, the only coupling for which an evidence is found, appears in the amplitudes only an overall normalization factor, with both $F_L^{\ell}$ and $A_{\rm FB}^\ell$ being insensitive to it: its effect is therefore only visible in $R(D^*)$. The reason why it is different from 0 is due to the fact that all the remaining couplings, being $m_\ell$ suppressed in interference terms of scalar or tensor operators with the SM contribution, are not capable to affect the predictions of $F_L^{\ell}$ and $A_{\rm FB}^\ell$. Therefore, in order to reproduce these observables, similarly to the SM case the FFs have to be stretched assuming values similar to the ones of the IgWa or BGJS approaches, where an anomaly in $F_L^{\ell}$ and $A_{\rm FB}^\ell$ is not present, at the cost of a reemerging one in the LFUV ratio: therefore, in order to address this last deviation, a non-vanishing value for $C_{V_L}$ is required.

To summarize, we observed that the prediction for $F_L^{\ell}$ and $A_{\rm FB}^\ell$ are crucial quantities in the estimate of FFs, due to their extremely feeble dependence on NP contributions contrarily to $R(D^*)$. We therefore conclude that any present and future determination of FFs in $B \to D^*$ transitions have to necessarily confront themselves with data on $F_L^{\ell}$ and $A_{\rm FB}^\ell$ with light leptons: indeed, while a discrepancy in $R(D^*)$ can be ascertained to NP effects, this is not the case for the former quantities.


\vspace{1em}
{\bf Acknowledgements.} This research was supported by the Deutsche Forschungsgemeinschaft (DFG, German Research Foundation) under grant  396021762 - TRR 257, from the Generalitat Valenciana (Grant PROMETEO/2021/071) and by MCIN/AEI/10.13039/501100011033 (Grant No. PID2020-114473GB-I00). M.F. wishes to thank M. Blanke, A. Crivellin, S. Iguro, T. Kitahara, U. Nierste, S. Simula, L. Vittorio and R. Watanabe, co-authors of the original papers.

\bibliographystyle{JHEP-CONF}
\bibliography{BIB}

\end{document}